\documentclass[journal]{IEEEtran}
\usepackage{amsmath,amsfonts,amssymb,amsthm,hyperref,color,graphicx}
\usepackage{enumerate}
\usepackage{newlfont}
\usepackage{caption}
\usepackage{subcaption}
\usepackage{epstopdf}
\usepackage{textcomp}
\usepackage{multirow}
\usepackage{algorithm}
\usepackage{algpseudocode}
\usepackage{url}
\usepackage{ragged2e}
\usepackage{ifpdf}
\usepackage{flushend}

\begin{document}
% \nocite{*}
\title{A study and comparison of COordinate Rotation DIgital Computer (CORDIC) architectures}
% \title{Marking the $64^{th}$ year of CORDIC}

\author{Neha K. Nawandar$^1$ and Vishal R. Satpute$^2$ \\{~\IEEEmembership{$^1$Department of Electrical Engineering, IIT Delhi and $^2$Department of ECE, VNIT, Nagpur}}}

\maketitle

%-----------------------------------------------------------------------------------------------------------------------

\begin{abstract}
Most of the digital signal processing applications performs operations like multiplication, addition, square-root calculation, solving linear equations etc. The physical implementation of these operations consumes a lot of hardware and, software implementation consumes large memory. Even if they are implemented in hardware, they do not provide high speed, and due to this reason, even today the software implementation dominates hardware. For realizing operations from basic to very complex ones with less hardware, a Co-ordinate Rotation Digital Computer (CORDIC) proves beneficial. It is capable of performing mathematical operations right from addition to highly complex functions with the help of arithmetic unit and shifters only. This paper gives a brief overview of various existing CORDIC architectures, their working principle, application domain and a comparison of these architectures. Different designs are available as per the target, i.e. high accuracy and precision, low area, low latency, hardware efficient, low power, reconfigurability, etc. that can be used as per the application in which the architecture needs to be employed.
\end{abstract}


%-----------------------------------------------------------------------------------------------------------------------

\IEEEpeerreviewmaketitle

%-----------------------------------------------------------------------------------------------------------------------

\section{Introduction} \label{introduction}
For years long the digital signal processing applications have been dominated by microprocessors. Though these processors are low cost and flexible enough, they are often not that fast for certain DSP tasks. The reconfigurable logic in computers permits higher speed at the same cost as that of the traditional software approach. Unfortunately algorithms for these microprocessors based systems are not totally compatible with the hardware. Hardware efficient solutions exist but software systems are still dominating the market and have kept these solutions out of the spotlight.  In the hardware efficient algorithms, there exists a class of iterative solutions that use only shifts and adds to perform trigonometric and other transcendental functions. This class is called as CORDIC which is an acronym for Coordinate Rotation Digital Computer as stated in \cite{volder1959cordic}, \cite{volder2000birth} which is used for DSP applications \cite{hu1992cordic}, \cite{haviland1980cordic}, for fast VLSI implementation \cite{duprat1993cordic} and for FPGA designs \cite{valls2002evaluation}, \cite{andraka1998survey}.

We know that, there is a need to maintain a trade-off between speed, power, area and accuracy, as benefits in all the design metrics cannot be achieved at the same time. This leads to the demand of approximate designs for error tolerant applications. High power and energy savings can be achieved by introducing tolerable errors in the architecture, which in turn helps to improve the design and quality metrics. We know that DCT is one of the most compute intensive blocks in DSP applications, which demands for an architecture that is capable of reducing its complexity. One such architecture is given in \cite{Nawandar2016} that proposes a low power CORDIC architecture and a low power approximate DCT architecture using the same. Such approximate architectures can be effectively used for DSP applications involving human intervention by efficiently exploiting persistence of vision. Some CORDIC based DCT architectures are also discussed in \cite{leavline2014cordic}, \cite{prasannan2014cordic}, \cite{xiao2012novel}, \cite{wang1994systolic}, \cite{sung2006high}, \cite{sun2007low}, \cite{chi2006low} and \cite{hu1995efficient}. A reconfigurable low power DCT architecture based on data priority is proposed by Lee et al. in \cite{lee2014reconfigurable}. 

\begin{figure*}
 \centering
 \includegraphics[height = 5.5cm, width = 1.5\columnwidth]{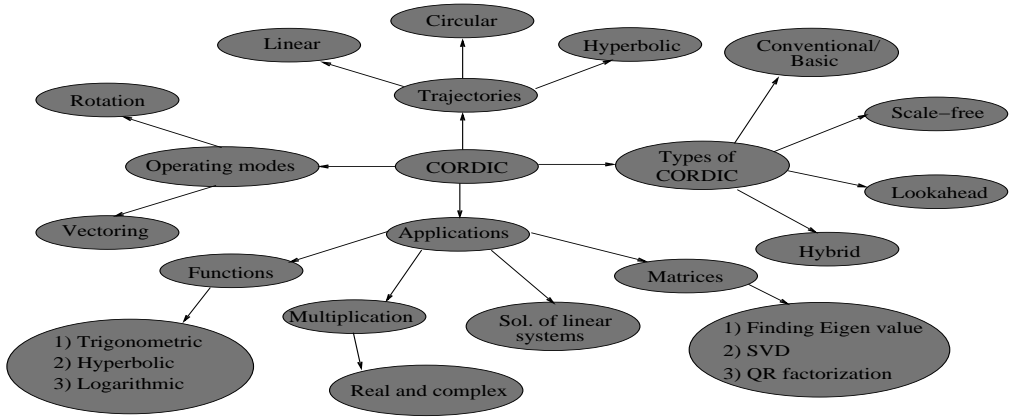}
 \caption{CORDIC: An overview}
 \label{cordictree}
\end{figure*}

The remaining paper is organized as follows: Section \ref{discussioncordic} explores CORDIC, gives its working principle and different modes, trajectories for which it can work. It also mentions a broad overview of the work covered in this papers (shown in Figure \ref{cordictree}). The different existing CORDIC architectures are mentioned in Section \ref{literature}. Results and discussions are made in Section \ref{discussioncordic}, and the paper is concluded in Section \ref{conclusion}. 

%-----------------------------------------------------------------------------------------------------------------------

\section{Exploring CORDIC and its application domain} \label{discussioncordic}
As mentioned in section \ref{introduction}, COrdinate Rotation DIgital Computer i.e. CORDIC is nothing but an easy way of performing simple to very complex mathematical computations. It can very efficiently perform several computing tasks like, calculation of trigonometric, functions, real and complex multiplications, division, multiplication, square-roots, cube-roots, finding solutions of linear equations, singular value decomposition, etc. The novelty of CORDIC is that it obtains the results using minimum hardware and using only adders/ subtracters, which consume very little area on a chip. So it would not be wrong to state that, it is a less complex and hardware efficient way of performing mathematical computing, applications in image, video processing, robotics, communication systems, etc.

\subsection{The CORDIC Algorithm}
The CORDIC algorithm was first introduced in 1959 by Jack E. Volder. It was developed at the aero-electronics department of Convair to replace the analog resolver in the B-58 bomber's navigation computer. It has been used in applications like 8087 co-processor, pocket calculators, radar signal processors, robotics etc. CORDIC is nothing but a computer which contains a special serial arithmetic unit consisting of shift and addition units, and special interconnections. By using a prescribed sequence of conditional add/subtract operations, the CORDIC arithmetic unit can be controlled and thus used to solve any function.

The main idea behind the working of CORDIC lies in the fact that, the co-ordinates of any vector are nothing but trigonometric returns of the angle it makes with the positive X-axis (cosine and sine values). So, whenever any vector is rotated in space, along with the new co-ordinates, it provides the angle parameters. This concept is the backbone of CORDIC algorithm. The input and output vectors are related with each other by a rotation matrix called as \emph{Givens rotation}. It helps to find out the final co-ordinates of the rotated vector.

\subsubsection{Givens Rotation}
It relates the input and output vectors and the relation between the initial and final vector is represented by \emph{Givens rotation} \cite{givens1958computation}, a rotation in the plane spanned by two coordinates 
axes. Its general form is,

\begin{equation}
\left[ \begin{array}{ccccccc}
1 & \cdots & 0 & \cdots  & 0 & \cdots & 0 \\
\vdots & \ddots & \vdots &  & \vdots &  & \vdots \\
0 & \cdots & c & \cdots  & -s & \cdots & 0 \\
\vdots &  & \vdots & \ddots & \vdots &  & \vdots \\
0 & \cdots & s & \cdots  & c & \cdots & 0 \\
\vdots & & \vdots & & \vdots &  \ddots &   \vdots \\
0 & \cdots & 0 & \cdots  & 0 & \cdots & 1
\end{array}\right]
\end{equation}
	
The following matrix works in case of 2D,

\begin{equation}
\left[ \begin{array}{cc}
cos\theta & -sin\theta \\
sin\theta & cos\theta
\end{array}\right]
\label{eq:rotmatrix}
\end{equation}

\begin{figure}[!t]
\centering \includegraphics[scale=0.8]{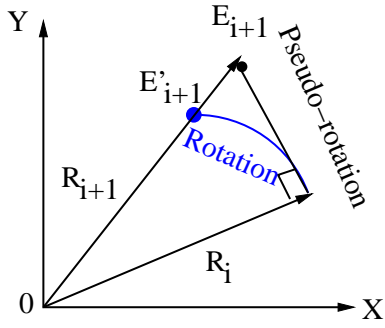}
\caption{Pseudo-rotation and real rotation \cite{meher200950}}
\label{fig:xyaxis}
\end{figure}

When we rotate a vector, it does not move linearly, instead it follows a circular path. Finding out the co-ordinates, when a linear path is followed involves the direct use of Givens rotation and some simple calculations. But for the real rotations involved here, finding the values is somewhat tricky, which is clear from Figure \ref{fig:xyaxis}. So, to easily find out the end points of finally obtained output vector, instead of considering the real rotation, pseudo-rotations are used. These rotations use the Pythagoras theorem to return the co-ordinate values and by further using them, the actual ones are found out.
	
This helps to reduce the architecture complexity but at the same time introduces an extra term/constant to compensate for the error occurring due to the pseudo-rotation. This term is often referred to as \emph{scaling factor}.

\subsubsection{Micro-rotations}
\begin{figure}[!t]
\centering \includegraphics[scale=0.8]{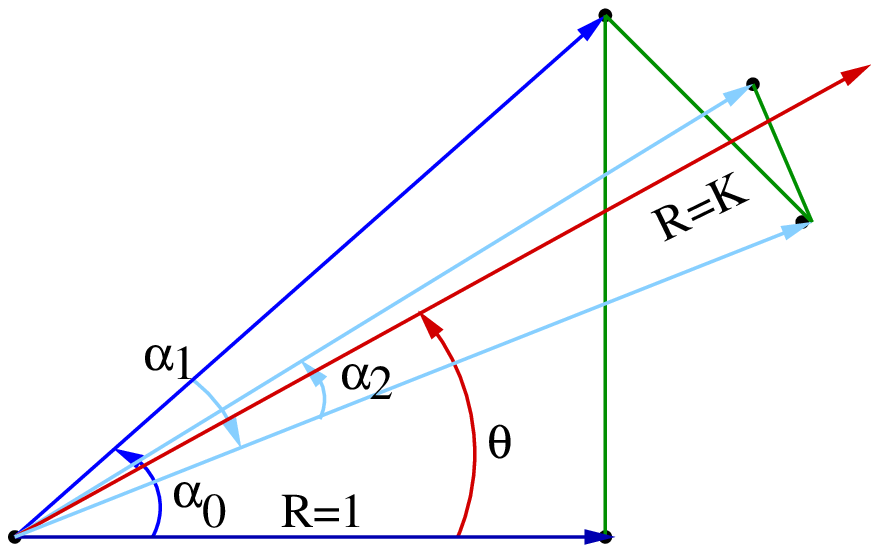}
\caption{Micro-rotations instead of one single large rotation}
\label{fig:rotation}
\end{figure}

For achieving an angular shift, CORDIC breaks it into a number of small angles as shown in Figure \ref{fig:rotation} and instead of rotating the initial vector by one large angle, it is rotated number of times by the broken angles. This can be called as iterative decomposition of angle of rotation. By doing this, there 
is ease in rotation and reduction in complexity of the algorithm. The steps wherein every time the vector is rotated are termed as \emph{iterations} which provide easy understanding of the algorithm. But increasing the iterations above an extent will add to its complexity and hardware cost.

Number of iterations can be limited by, either fixing it or comparison at each stage with the residual angle. These are generally referred as \emph{iterations} and denoted by \emph{i}. In CORDIC algorithm, to simplify 
the micro-rotations, instead of using the direct inverse tangent values, divide by increasing powers of 2 is done. It is also necessary to decide which way the vector is to be rotated i.e. in clockwise or in anti-clockwise direction. This will be known using the value of $\sigma$, which depends upon the sign of residual angle.

\subsubsection{The CORDIC Gain}
Whenever vector is normally rotated, the co-ordinates it gives are the accurate ones. But when a real rotation is replaced by a pseudo-rotation it gives erroneous co-ordinate values. This is because, whenever a vector is rotated, as that we do here; the length of final vector is more than the actual required one which we would have obtained by real-rotation. The factor by which the original accurate length differs from the obtained one 
is termed as \emph{scaling factor (k)} \cite{meher200950}, and is given by the following equation,

\begin{equation}
k_{i}=\dfrac{1}{\sqrt{1+2^{-2i}}}
\label{sf}		
\end{equation}
		
It provides the scaling factor value for a particular value of \emph{i}. \emph{k} is independent of the direction of micro-rotations and it decreases monotonically and finally converges to 0.6705 as number of rotations tends to $\infty$. The inverse of \emph{k} is called as CORDIC Gain. The exact scaling-factor value \cite{meher200950} depends on the number of iterations and is given by the following equation,

\begin{equation}
k_{n}=\prod_{n}\dfrac{1}{\sqrt{1+2^{-2i}}}
\end{equation}
		
Redundant CORDIC methods with a constant scale factor for sine and cosine computation is mentioned in \cite{takagi1991redundant}. The above mentioned equations form the backbone of CORDIC algorithm. These will be used in most of the architectures and are mentioned in the forthcoming sections. The DSP applications today demand low power, error resiliency and high performance. To achieve this, there is a need to design architectures capable of addressing these issues. Most of the image or video processing applications constitute DCT in its architecture, which is considered as the most power hungry block. This research work considers calculations of the DCT matrix coefficients by employing CORDIC algorithm to ensure reduced hardware and software complexity.

CORDIC is used in its rotation mode of operation to generate the \emph{cosine} values of some specific angles, which are nothing but the DCT matrix coefficients and can be found out using various CORDIC architectures. In a DCT matrix of particular dimension the value of angles for which the coefficients need to be found out is fixed. Thus, in this case, \emph{angle recoding scheme} can be used, as it is efficient only if we previously know the value of angles for which the cosine values are to be generated.

Scaling factor also plays an important role in determining the latency and complexity of operation. A scale-free design is thus to be exploited, so that the design metrics are improved. As the image processing applications involve human intervention, persistence of vision can be exploited so as to get nearly accurate results, which means that approximation is to be done.  It is clear that CORDIC uses adders in its hardware, so instead of using accurate adders, approximate adders can be utilized which will consume less power and provide with tolerable errors in the results. 

The concept of hybrid angles is also to be employed as it would limit the iterative nature of CORDIC to certain extent and introduce partial parallelism. To reduce the number of iterations and avoid data dependency, \emph{lookahead approach} proves beneficial, as, it is capable of performing more than one rotation in a single iteration. The above mentioned concepts will be used to propose a novel error tolerant and energy efficient CORDIC architecture.

%-----------------------------------------------------------------------------------------------------------------------

\section{Literature survey} \label{literature}
This section covers various CORDIC architectures and the advantages and limitations associated with them that have been studied during the literature review.

\subsection{Conventional CORDIC architecture}
Proposed in the year 1959, \cite{volder1959cordic} is the very first CORDIC architecture, which talks about the basic architecture of CORDIC, the input and output relation, rotations, the gain, modes of operation etc. which are also discussed here. The input and output vectors are related by Givens Rotation  \cite{givens1958computation} as stated in equation. \ref{gr} and it is modified to equation \ref{gr1},

\begin{equation}
\left[ \begin{array}{c}
x_{i+1}\\
y_{i+1}
\end{array}\right]=
\left[ \begin{array}{cc}
cos\theta & -sin\theta\\
sin\theta & cos\theta
\end{array}\right]
\left[ \begin{array}{c}
x_i\\
y_i
\end{array}\right]
\label{gr}
\end{equation}  	
	
\begin{equation}
\left[ \begin{array}{c}
v_{i+1}
\end{array}\right]=
\left[ \begin{array}{cc}
cos\theta & -sin\theta\\
sin\theta & cos\theta
\end{array}\right]
\left[ \begin{array}{c}
v_{i}
\end{array}\right]
\label{gr1}
\end{equation}
	        
\noindent where, [v$_{i+1}$] is the vector at (i+1)$^{th}$ iteration and [v$_{i}$] is the vector at i$^{th}$ iteration.
Let the Givens Rotation matrix be denoted by R$_{g}$. R$_{g}$ is considered for modifications and $\frac{1}{cos\theta}$ is taken out as a common factor, which reduces the matrix to,

\begin{equation}		  		  	
\left[ \begin{array}{cc}
1 & -tan\theta \\
tan\theta & 1
\end{array}\right]
\end{equation}	
	
\begin{figure}[!t]
\centering
\includegraphics[height=6cm, width=\columnwidth]{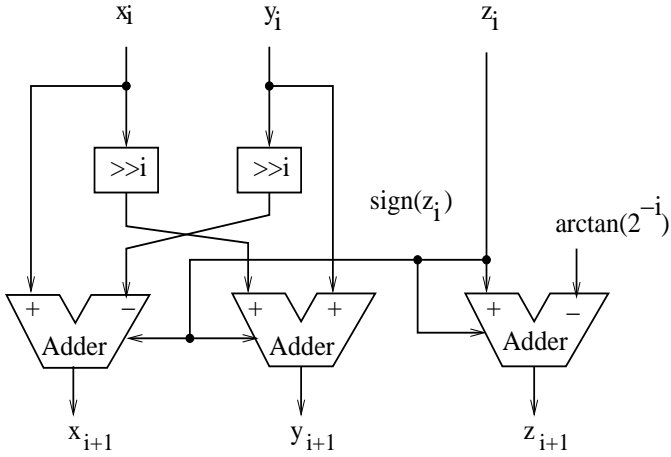}
\caption{Basic CORDIC Architecture \cite{meher200950}}
\label{fig:conv}
\end{figure}

For calculations in conventional CORDIC this matrix is always considered, as it is convenient to represent tan$\theta$ terms of 2$^{-i}$. But it also brings a disadvantage with it, i.e the $\frac{1}{cos\theta}$ term which is nothing but the scaling factor as mentioned above. Applying trigonometric rules, the term gets modified to equation \ref{sf}. Figure \ref{fig:conv} shows the basic CORDIC \cite{volder1959cordic} architecture. 

The introduction of scaling-factor simplifies the algorithm by saving on angle calculations, but at the same time it introduces the need of multiplier to correct the final vector length/ co-ordinate values. Though it incurs more hardware, we cannot totally neglect it as, it introduces non-tolerable error. Approaches to deal with this issue have been explained further.

To understand the working of CORDIC it is primarily required to know about its mode of operation. CORDIC basically operates in two modes for three trajectories i.e. linear, circular and hyperbolic, depending upon the operation required to perform. The modes are, (1) Rotation Mode and (2) Vectoring Mode. The conventional architecture is given by Figure \ref{fig:conv} and the basic equations on which CORDIC relies as mentioned in \cite{meher200950} are,

\begin{equation}
x_{i+1} = k(x_i cos\theta_i - y_i sin\theta_i) 
= x_i - \sigma_i2^{-i}y_i 
\end{equation}				

\begin{equation}
y_{i+1} = k(y_i cos\theta_i + x_i sin\theta_i) 
= y_i + \sigma_i2^{-i}x_i
\end{equation}				

\begin{equation}			
z_{i+1} = z_{i} - \sigma_{i} tan^{-1}(2^{-i})
\end{equation}
				
where, z$_{i+1}$ is the remaining value of angle after i$^{th}$ iteration i.e. the residual angle.

\subsubsection{Rotation Mode:} In this mode of operation, the input vector is rotated until the residual angle becomes zero. After every rotation is performed, the residual angle is checked for both its sign and value as, the sign tells about the direction of next rotation and value ensures whether another iteration needs to be performed or not. Generally, the initial values of x and y co-ordinates are taken as, 1 and 0 respectively, whereas, z will contain the angle by which a rotation needs to be performed. The equations \ref{rotmodex} and \ref{rotmodey} are performed until equation \ref{rotmode} is satisfied.

\begin{equation}
x_{n} = k(x_0 cos\theta_0 - y_0 sin\theta_0)
\label{rotmodex}
\end{equation}
				  
\begin{equation}						
y_{n} = k(y_0 cos\theta_0 + x_0 sin\theta_0)
\label{rotmodey}
\end{equation}
				  
\begin{equation}	
z_{n} = 0
\label{rotmode}
\end{equation}

\begin{figure*}[!t]
    \centering
    \begin{subfigure}[b]{0.3\textwidth}
        \centering
        \includegraphics[width=\textwidth]{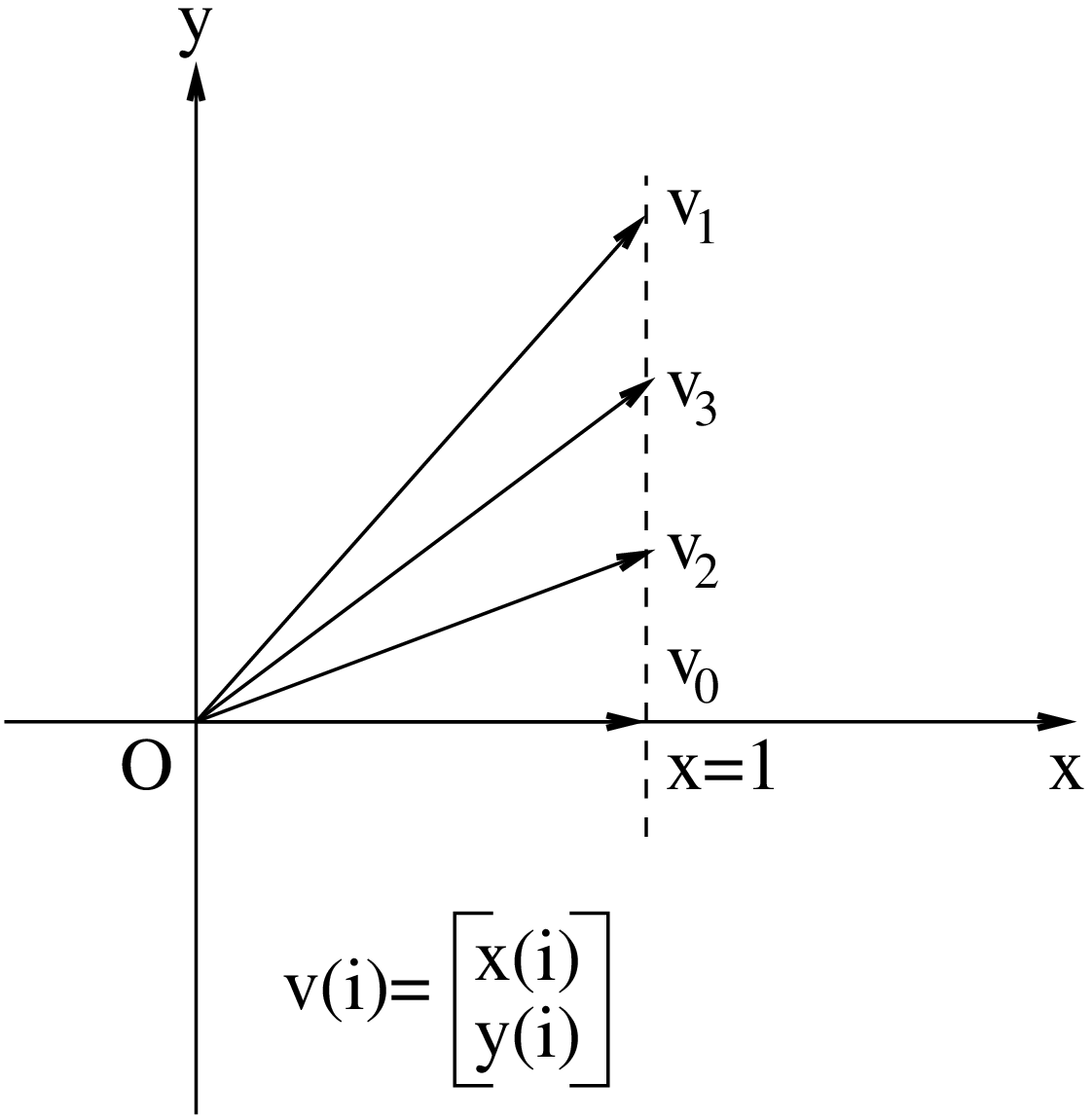}
        \caption{Linear}
         \label{fig:traj1}
    \end{subfigure}
    \hfill
    \begin{subfigure}[b]{0.3\textwidth}
        \centering
        \includegraphics[width=\textwidth]{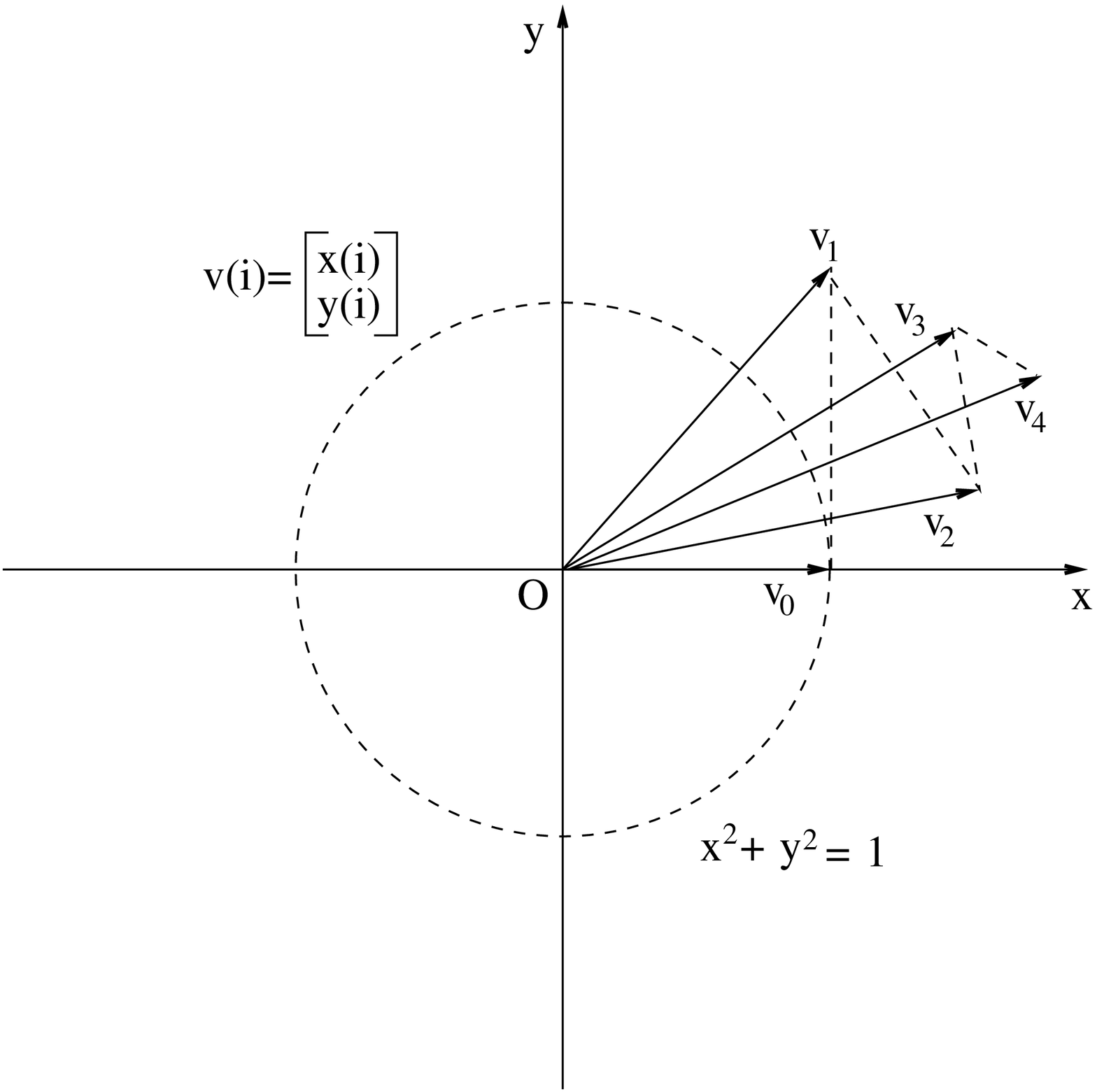}
       \caption{Circular}
       \label{fig:traj2}
    \end{subfigure}
    \hfill
    \begin{subfigure}[b]{0.3\textwidth}
        \centering
        \includegraphics[width=\textwidth]{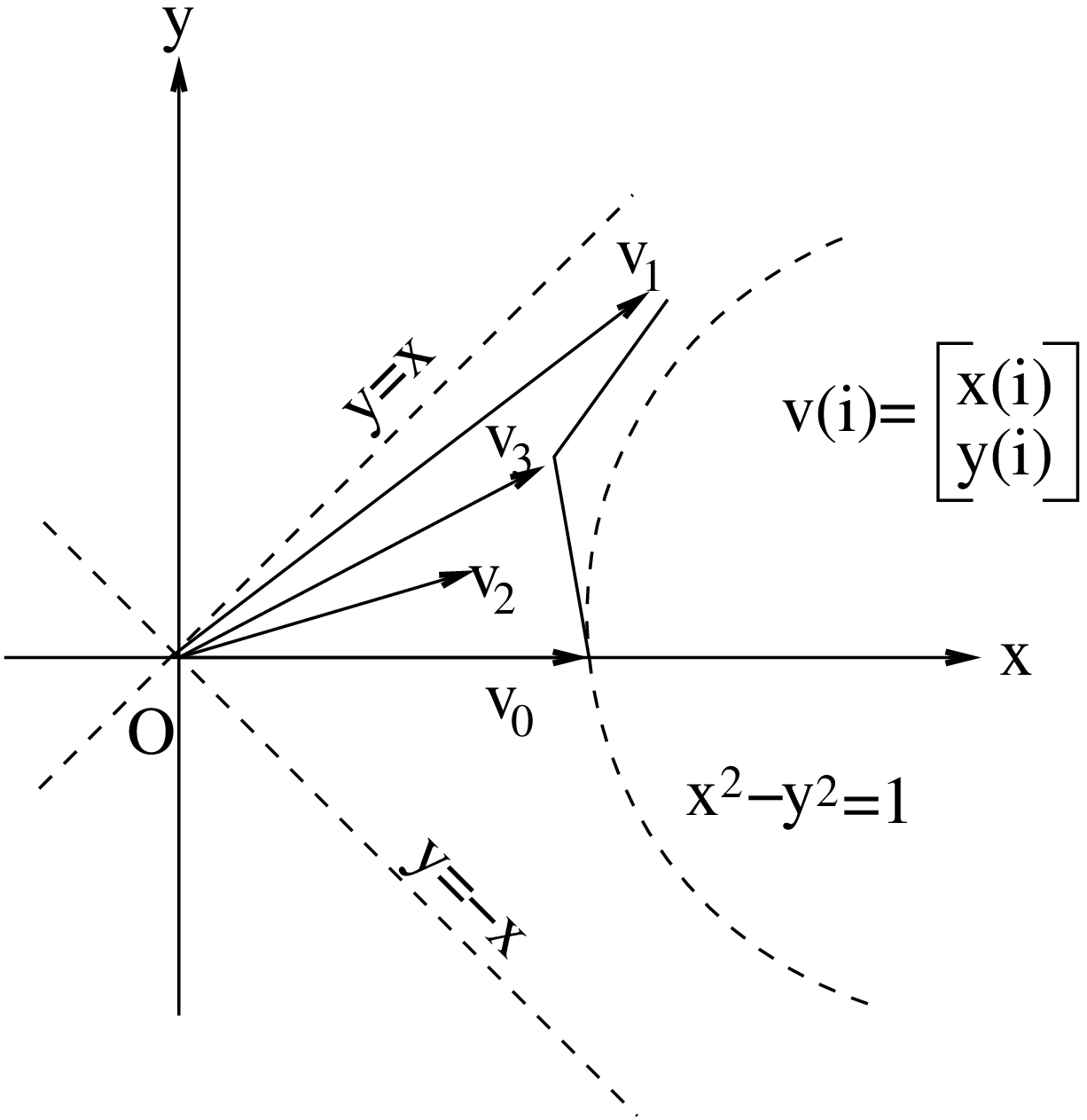}
        \caption{Hyperbolic}
        \label{fig:traj3}
    \end{subfigure}
    \caption{Different trajectories of CORDIC}
    \label{fig:traj}
\end{figure*}

\subsubsection{Vectoring Mode :} Here, the input vector is rotated until its y component gets reduced to zero i.e. the purpose is to take the vector near x-axis, in order to obtain the vector amplitude and phase. z$_{n}$ will contain the value by which the vector was rotated which in case of rotation mode was used as a residual angle storing register. The equations for this mode, as mentioned in \cite{meher200950} are given below,

\begin{equation}
x_{n}=k\sqrt{{x_{0}^{2}+y_{0}^{2}}}
\end{equation}
		
\begin{equation}
y_{n}=0
\end{equation}
		
\begin{equation}
z_{n}=z_{0}+tan^{-1}(\dfrac{y_{0}}{x_{0}})
\end{equation}

CORDIC is an algorithm which allows operation in different modes and trajectories as needed for e.g. for normal trigonometric functions it operates in circular mode while for the hyperbolic ones it will work in hyperbolic trajectory. Figure \ref{fig:traj} as mentioned in \cite{lakshmi2010cordic} shows rotation mode of CORDIC for all the three trajectories. A reconfigurable architecture which considers both modes for circular and hyperbolic trajectories is mentioned in \cite{aggarwal2014reconfigurable}.

Certain limitations of conventional CORDIC architecture are,
\begin{enumerate}[i]
\item Need of scale-factor calculation.
\item Increased latency and data dependency due to iterative nature.
\item Selection of direction of micro-rotations.
\end{enumerate}

\subsection{Hardware efficient CORDIC architecture (Scale-free CORDIC)}
\begin{figure}[!t]
 \centering
 \includegraphics[height = 6cm, width = \columnwidth]{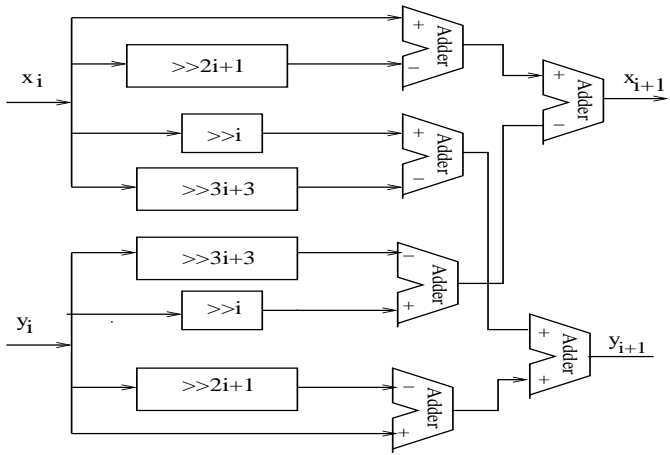}
 \caption{Co-ordinate computation \cite{aggarwal2012hardware}}
 \label{scalefree}
\end{figure}

Scale-factor calculation requires a multiplier which incurs extra hardware. But it cannot be totally neglected, as it leads to an intolerable error of $\sim$ 60 \%. Instead of totally neglecting it, an alternative to scale-factor, such that no loss of accuracy is there, would be a much better solution. Design of scale-free CORDIC, as mentioned in \cite{aggarwal2012hardware}, \cite{causo2012parallel}, \cite{aggarwal2012area} is divided in two steps, first one is the co-ordinate calculation (shown in Figure \ref{scalefree}) while the second step is the micro-rotation sequence identification (Table \ref{lobd}). A new efficient window-architecture design using completely scaling-free CORDIC pipeline is discussed in \cite{aggarwal2013efficient} and a scale-free hyperbolic CORDIC processor is mentioned in \cite{aggarwal2013scale}.

The Rotation matrix which is given by equation \ref{eq:rotmatrix}, contains sine and cosine terms, which on simplification introduces scaling-factor. If these terms are replaced by their respective Taylors' series, there would not be any requirement for finding out the scaling factor. Approximation of third, fourth or any order, applied to Taylor series will replace the complex sine and cosine terms by simple terms that can be implemented by simple shift operations. The rotation matrix then gets reduced to,

\begin{equation}
\left[ \begin{array}{cc}
1-\theta^{2}/2 & -\theta+\theta^{3}/6 \\
\theta-\theta^{3}/6 & 1-\theta^{2}/2
\end{array}\right]	
\label{sfc1}
\end{equation}
		
It is obtained after applying third order of approximation to the Taylor series of \emph{sine} and \emph{cosine} functions. This helps to find out the next iteration values and is used in the co-ordinate calculation unit. For identifying the micro-rotations it uses \emph{leading-one bit detection technique}. The main idea of scale-free CORDIC is restricting the micro-rotations to anticlockwise direction only. The leading-one bit detector is discussed below.

Leading-one bit detector follows the following steps:
\begin{enumerate}[i]
\item Firstly the angle to be rotated is converted into binary and then into hexadecimal format.
\item Then the leading one bit is found out, calculating from the MSB bit position i.e. starting from count 15 and it is made 0.
\item This is continued up to the point where the value (angle z) gets reduced to 0.
\end{enumerate}

Certain limitations of scale-free CORDIC are stated below:
\begin{enumerate}[i]
\item The region of convergence (ROC) provided by scale-free CORDIC is very less and thus it is less suitable for practical applications.
\item Because of the approximation applied to the Taylor series, there is a restriction on the start iteration value index \emph{i}. 
% This is as per the equation,
% $$i=\left\lfloor{\{b-\log_{2}(4!)\}\over 4}\right\rfloor, {\rm where} \ b \ 
% {\rm is} \ {\rm the} \ {\rm wordlength}\eqno{\hbox{(8)}}$$
\item It is best suitable only for 16 bit data word length, because if we increase the bit width, the number of iterations increase, thus leading to more complexity.
\end{enumerate}

\begin{table}[!t]
\centering
\caption{Leading-one bit detector operation for an angle of $27^\circ$}
\vspace{8pt}
\begin{tabular}{p{0.8cm}p{0.8cm}p{1.5cm}p{1.8cm}p{2cm}}
 \hline \noalign {\smallskip}
 Stage & z$_i$ & Lead One Position (LOP) & Shift (i = 16-LOP) & z$_{i+1}$ \\
 \noalign {\smallskip} \hline \noalign {\smallskip}
 1 & 78A3 & 15 & 2  & 78A3-4000 = 38A3 \\
 2 & 38A3 & 13 & 3  & 18A3 \\
 3 & 18A3 & 12 & 4  & 08A3 \\
 4 & 08A3 & 11 & 5  & 00A3 \\
 5 & 00A3 &  7 & 9  & 0023 \\
 6 & 0023 &  5 & 11 & 0003 \\
 7 & 0003 &  1 & 15 & 0001 \\
 \noalign {\smallskip} \hline
 \end{tabular}
 \label{lobd}
 \end{table}

CORDIC in general is better in terms of hardware requirement, but it consumes more time to generate the output as compared to other methods. So, to overcome this limitation, some CORDIC architectures that have less latency as given in \cite{shukla2012low}, \cite{wang1997hybrid} are discussed further.

\subsection{Lookahead CORDIC architecture}
Basic CORDIC is iterative in nature which introduces data dependency due to the following reasons,
\begin{enumerate}
\item Next micro-rotation can be done only if the previous rotation has been completed.
\item It is performed on the vector which is obtained by the previous rotation.
\end{enumerate}

Thus, it becomes very difficult to remove data dependency. The only solution to this is, an architecture as discussed in \cite{lee2014reconfigurable} which previously detects the direction of micro-rotation so that more number of these can be performed in a single iteration. This reduces both, the number of iterations, thus indirectly reducing the latency and data dependency.

\begin{figure}
\centering
\includegraphics[width=0.75\columnwidth]{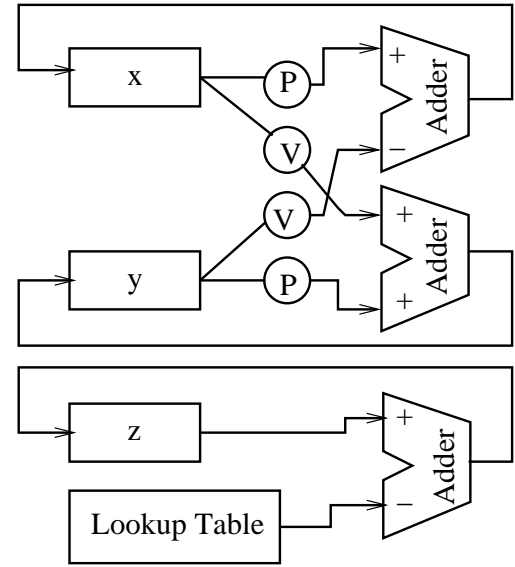}
\caption{Block diagram of Lookahead CORDIC}
\label{fig:lac}
\end{figure}

One such architecture which employs this concept is \emph{Lookahead CORDIC}. Figure \ref{fig:lac} shows the block diagram of Lookahead CORDIC. The principle of working is similar to that of a carry lookahead adder. Lookahead means that a number of CORDIC iterations can be computed ahead to finish the iterations at one time. The basic idea of the proposed   scheme is to reduce the iteration number directly while maintaining the precision. The design consists of vertical and parallel circuits which help to remove the data dependency. This approach has very low latency which reduces when more iteration are done within a single rotation.  Following mentioned are the equations for four consecutive iterations.

\begin{equation}
x_1=x_0-\sigma_0 2^{-0}y_0
\end{equation}

\begin{equation}
y_1=y_0+\sigma_0 2^{-0}x_0
\end{equation}

\begin{equation}
x_2=x_1-\sigma_1 2^{-1}y_1
\end{equation}

\begin{equation}
y_2=y_1+\sigma_1 2^{-1}x_1
\end{equation}

\begin{equation}
x_3=x_2-\sigma_2 2^{-2}y_2
\end{equation}

\begin{equation}
y_3=y_2+\sigma_2 2^{-2}x_2
\end{equation}

\begin{equation}
x_4=x_3-\sigma_3 2^{-3}y_3
\end{equation}

\begin{equation}
y_4=y_3+\sigma_3 2^{-3}x_3
\end{equation}

These equations will be then used to find out the value of $\sigma_0$, $\sigma_1$, $\sigma_2$ and $\sigma_3$, which will thus avoid the requirement of more number of iterations, as, number of rotations will be performed in a single iteration. This also eliminates the data dependency which was relevant in the conventional CORDIC architecture, as it was iterative in nature. It is clear from the equations mentioned below. These are obtained by substituting the values of $x_1$, $x_2$, $x_3$ and $y_1$, $y_2$, $y_3$ in the above equations, so as to get a relation between $x_0$ and $x_4$ \& $y_0$ and $y_4$. The final equations obtained are,

\begin{equation}
x_4=x_0P-y_0V
\end{equation}

\begin{equation}
y_4=y_0P+x_0V
\end{equation}

where,

\begin{equation}
P=
\left[ \begin{array}{c}
2^{-0}(+1)\\
2^{-1}(-\sigma_0\sigma_1)\\
2^{-2}(-\sigma_0\sigma_2)\\
2^{-3}(-\sigma_1\sigma_2-\sigma_0\sigma_3)\\
2^{-4}(-\sigma_1\sigma_3)\\
2^{-5}(-\sigma_2\sigma_3)\\
2^{-6}(+\sigma_0\sigma_1\sigma_2\sigma_3)	
\end{array}\right]	
\end{equation}
	      	
\begin{equation}
V=
\left[ \begin{array}{c}
2^{-0}(+\sigma_0)\\
2^{-1}(+\sigma_1)\\
2^{-2}(+\sigma_2)\\
2^{-3}(-\sigma_1\sigma_2\sigma_0+\sigma_3)\\
2^{-4}(-\sigma_0\sigma_1\sigma_3)\\
2^{-5}(-\sigma_0\sigma_2\sigma_3)\\
2^{-6}(-\sigma_1\sigma_2\sigma_3)	
\end{array}\right]
\end{equation}
	
These parallel shift (P) and vertical shift (V) circuits are then generated, which are finally used to get the required output values. The conventional CORDIC is iterative in nature and so, the number of iterations in the two data paths of same CORDIC cannot be changed. When there is a need to perform different number of iterations for the two data paths, for a single CORDIC, this approach is the solution. As here, the micro-rotation direction for more iterations is found out previously, data dependency is removed, and so the next iteration no more dependent on the previous ones. This allows performing more rotations in a single iteration, which thus reduces the latency. The \emph{drawback} it carries is that, it requires extra circuitry for developing 
the parallel and vertical shift logics, which increases with an increase in the number of iterations performed within a single rotation.

\subsection{Hybrid CORDIC architecture}
\begin{figure}[!h]
 \centering
 \includegraphics[height = 3.5cm, width = \columnwidth]{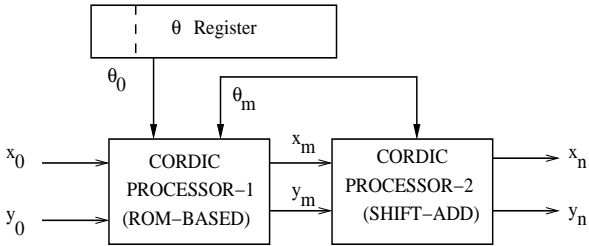}
 \caption{Mixed Hybrid CORDIC architecture \cite{wang1997hybrid}}
 \label{hybrid2}
\end{figure}

\begin{figure}[!t]
 \centering
 \includegraphics[height = 3.5cm, width = \columnwidth]{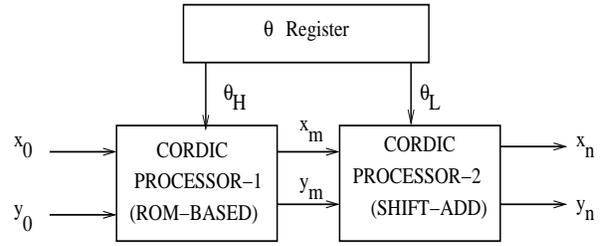}
 \caption{Partitioned Hybrid CORDIC architecture \cite{wang1997hybrid}}
 \label{hybrid1}
\end{figure}

The core idea behind CORDIC algorithm is that the angles are represented in terms of to the power of \emph{2}, thus it can be said that the initial angle is a sum of arc tangent constants. The vector will be rotated by these fixed angular values, the only difference being the direction in which it will be rotated. This depends upon the direction of micro-rotation i.e. $\sigma_i$, which makes it dependent thus making it difficult for parallelization. The hybrid CORDIC architectures \cite{wang1997hybrid} as shown in Figures \ref{hybrid2} and \ref{hybrid1} make it partially parallelized by computing $\sigma_i$ parallelly without affecting the accuracy of the algorithm.

The architecture of a mixed hybrid CORDIC consists of two cascaded CORDIC blocks, where the first block uses the input angle $\theta$ and the initial vector ($x_{in}$, $y_{in}$) to compute the residual angle $\theta_r$, the vector co-ordinates $x_{m}$, $y_{m}$ after some \emph{m} number of iterations. The second CORDIC block uses these inputs to obtain the final vector co-ordinates. This procedure leads to an increased complexity, but the throughput can be improved.

Another hybrid CORDIC architecture as shown in Figure \ref{hybrid1} is based on a partitioned algorithm where, the procedure is same as that of the mixed hybrid, the only difference being of giving the input angle. Here, the input angle is divided into MSB and LSB parts, and then further processing is done.

\subsection{Angle recoding CORDIC architecture}
Performing iterations until required result is obtained induces complexity if it increases beyond a limit i.e. if we get the necessary accurate output in less than 5 iterations then it is feasible and the hardware is less complex, but if increased beyond this limit, the hardware complexity increases. An architecture which reduces this uses the concept of \emph{angle recoding} \cite{hu1993angle}. The purpose of angle recoding (AR) is to reduce the number of CORDIC iterations by encoding the angle of rotation as a linear combination of a set of selected 
elementary angles of micro-rotations.

For applications where the angle of rotation is known in advance, a method to speed up the execution of the CORDIC algorithm by reducing the total number of iterations is presented. This is accomplished by using a technique called angle recoding, which encodes the desired rotation angle as a linear combination of very few elementary rotation angles. Each of these elementary rotation angles takes one CORDIC iteration to compute. The fewer the number of elementary 
rotation angles, the fewer the number of iterations are required.

A greedy algorithm which takes only O(n$^2$) operations is developed to perform CORDIC angle recoding. It is proven that this algorithm is able to reduce the total number of required elementary rotation angles by at least 50\% without affecting the computational accuracy. AR methods are well-suited for many signal processing and image processing applications where the rotation angle is known a priori, such as when performing the discrete orthogonal transforms like discrete Fourier transform (DFT), the discrete cosine transform (DCT), etc. Types of AR methods include,

\textbf{Elementary-Angle-Set recoding :} Using this recoding scheme the total number of iterations could be reduced by at least 50\% keeping the same n-bit accuracy unchanged. A similar method of angle recoding in vectoring mode called as the \emph{backward angle recoding}.

\textbf{Extended Elementary-Angle-Set Recoding (EEAS):} EEAS has better recoding efficiency in terms of the number of iterations and yields better error performance than the AR scheme based on EAS.

% \begin{figure}[here]
% \centering
% \includegraphics[scale=0.6, width=8cm]{eeas.png}
% \caption{Elementary-Angle-Set Recoding}
% \label{eeas}
% \end{figure}

\textbf{Parallel Angle recoding :} The AR methods could be used to reduce the number of iterations by more than 50\%, when the angle of rotation is known in advance. However, for unknown rotation angles, their hardware implementation involves more cycle time than the conventional implementation, which results in a reduction in overall efficacy of the algorithm.

\begin{figure}[!h]
\centering
\includegraphics[height=6cm, width=\columnwidth]{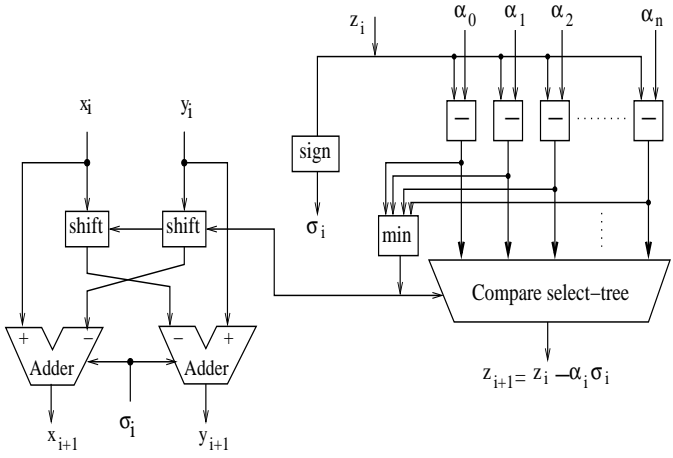}
\caption{Parallel Angle Recoding}
\label{parallel}
\end{figure}

To reduce the cycle time of CORDIC iterations in such cases, a parallel angle selection scheme can be used in conjunction with the AR method, to gain the advantages of the reduction in iteration count, without further increase in the cycle time. The elementary angles can be tested in parallel and the direction for the micro-rotations can be determined quickly to minimize the iteration period as shown in Figure \ref{parallel}.

CORDIC is iterative in nature, which causes data dependency between different iterations and also within the same iteration. If there is a need to perform different iterations for x and y coordinates, it is not possible using the conventional CORDIC, instead, an architecture which removes this data dependency is needed. The next subsection discusses about this in detail.

\subsection{Reconfigurable CORDIC architecture}
\begin{figure}[!h]
 \centering
 \includegraphics[height = 7cm, width = \columnwidth]{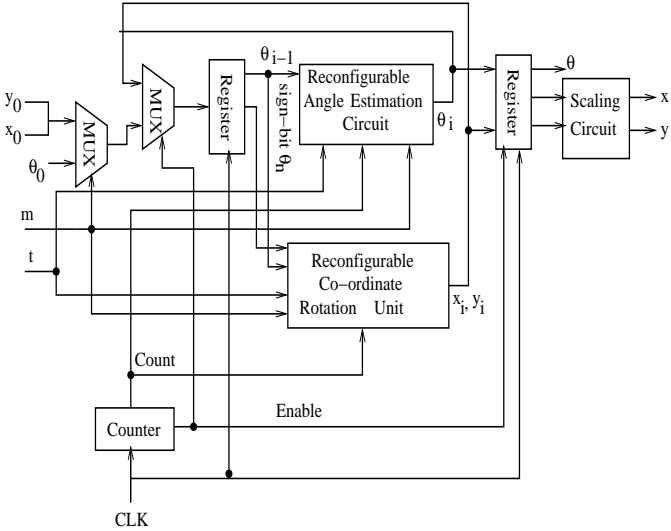}
 \caption{Reconfigurable CORDIC architecture}
 \label{reconfarchi}
\end{figure}

A reconfigurable CORDIC architecture has been presented by Aggarwal et al. in \cite{aggarwal2014reconfigurable} that can be configured to work for either of the circular or hyperbolic trajectories and in rotation and vectoring modes as well. The CORDIC is capable of computing various functions, trigonometric, exponential, logarithm, roots, etc. and Figure \ref{reconfarchi} shows the reconfigurable CORDIC architecture that can perform the mentioned functionality.

\subsection{Radix-4 CORDIC architecture}
This is a very old architecture, in which double shifts in each iteration, as compared to the conventional CORDIC are performed. This means that, instead of shifting the vector by \emph{i} bits in $i^{th}$ iteration, it is shifted by \emph{2i} bits in every $i^{th}$ iteration. In this way it helps in reducing the number of iteration
and leads to a significant reduction in the number of cycles for a word serial architecture which in turn reduces the latency. The equations get modified as,

\begin{equation}
x_{i+1}=x_{i}+\sigma_{i}4^{-i} y_{i}
\end{equation}

\begin{equation}
y_{i+1}=y_{i}-\sigma_{i}4^{-i} x_{i}
\end{equation}

\begin{equation}
z_{i+1}=z_{i}-\alpha_{i} \sigma_{i}
\end{equation}

Limitation of radix-4 CORDIC: The scaling factor in this design is not constant as in case of the conventional CORDIC. This is because, unlike in the basic CORDIC, where $\sigma$ carries values -1 and 1, here it carries the 
values, -2,-1, 1, 2, which restricts the scaling-factor from being a constant value. Thus, it is needed to find out the scaling factor which is more troublesome than conventional CORDIC. So, it incurs extra hardware and cycles to calculate the value of the scale factor. The equation of scaling factor here as mentioned in  \cite{antelo1997high}, is modified to equation \ref{r4}.

\begin{equation}
k=\prod_{i\geq0}k_{i}=\prod_{i\geqslant0}\sqrt{1+\sigma_{i}^{2}4^{-2i}}.
\label{r4}
\end{equation}

More number of iterations also increases the delay. So to overcome this limitation, an architecture which generates the output in less number of iterations, as mentioned next, is a solution.

\subsection{Repetitive Iteration CORDIC architecture} \label{ricosec}
\begin{figure}[!t]
  \centering
  \includegraphics[height=8cm, width=\columnwidth]{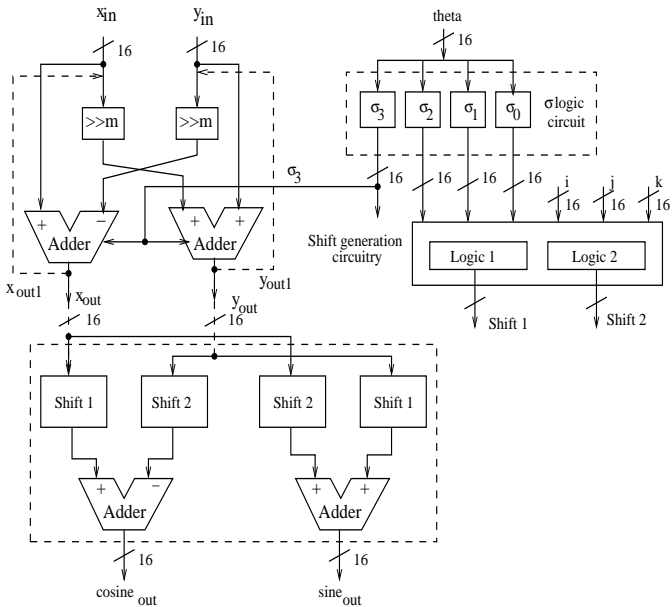}
  \caption{Repetitive Iteration CORDIC Architecture (RICO)}
  \label{Fig:RICO}
 \end{figure}

One of the major drawbacks of Conventional CORDIC is the data dependency due to its iterative nature. It is partially overcome by a repetitive iteration architecture (RICO) \cite{Nawandar2016} as shown in Figure \ref{Fig:RICO}. Here, the number of iterations that are to be performed is fixed beforehand and is the same for any value of input angle. The working of this architecture is broadly divided into two units: (i) The $\sigma$ generation unit and (ii) The shift generation unit.

For avoiding data dependency, the iteration number 0, 1 and 2 are performed in a single step, using a single CC. The coordinates of the vector that has to undergo through these three iterations is determined by rotating twice the initial input vector \textbf{IV($x_{in}$,$y_{in}$)}, where $x_{in}$ = 1 and $y_{in}$ = 0 by an angle of 7\textdegree. Coordinates of this vector i.e $OV$($x_{out}$, $y_{out}$) are used as input x and y values for the $0^{th}$, 
$1^{st}$ and $2^{nd}$ iteration. A quality tunable CORDIC with RICO as its core is presented in \cite{nawandar2018energy}.

 \begin{table*}
  \centering
  \caption{CORDIC algorithm for various computations}
  \label{cortable}
  \begin{tabular}{p{2.5cm}p{1cm}p{1cm}p{1cm}p{4cm}}
  
  \hline \noalign {\smallskip} 
   Required operation & Mode & Trajectory & Initial value & Final output \\
   \noalign {\smallskip} \hline \noalign {\smallskip}
   
   cos$\theta$, sin$\theta$, tan$\theta$ & Rotation & Circular & $x_{in}$ = 1, $y_{in}$ = 0, $\theta$ = $\theta_{in}$ & $x_{out}$ = cos$\theta$, $y_{out}$ = sin$\theta$ \\
   
   Polar to rectangular & Rotation & Circular & $x_{in}$ = R, $y_{in}$ = 1, $\theta$ = $\theta_{in}$ & x = R.cos$\theta$, y = R.sin$\theta$ \\
   
   cosh$\theta$, sinh$\theta$, tanh$\theta$ & Rotation & Hyperbolic & $x_{in}$ = 1, $y_{in}$ = 0, $\theta$ = $\theta_{in}$ & $x_{out}$ = cosh$\theta$, $y_{out}$ = sinh$\theta$ \\
   
   exp($\theta$) & Rotation & Hyperbolic & $x_{in}$ = 1, $y_{in}$ = 0, $\theta$ = $\theta_{in}$ & exp($\theta$) = cosh$\theta$ + sinh$\theta$ (using above) \\
   
   $tan^{-1}a$ & Vectoring & Circular & $x_{in}$ = a, $y_{in}$ = 1, $\theta$ = 0 & $\theta$ = $tan^{-1}a$ \\
   
   Rectangular to polar & Vectoring & Circular & $x_{in}$ = a, $y_{in}$ = b, $\theta$ = 0 & $x_{out}$ = $\sqrt{a^2 + b^2}$, $\theta$ = $tan^{-1}\frac{b}{a}$ \\
   
   Division $(\frac{b}{a})$ & Vectoring & Linear & $x_{in}$ = a, $y_{in}$ = b, $\theta$ = 0 & $\theta$ = $\frac{b}{a}$ \\
   
   ln(a), $\sqrt{a}$ & Vectoring & Hyperbolic & $x_{in}$ = a + 1, $y_{in}$ = a - 1, $\theta$ = 0 & $x_{out}$ = $\sqrt{a}$, $\theta$ = $\frac{1}{2}$ ln(a) \\
   
   \noalign {\smallskip} \hline 
  \end{tabular}
 \end{table*}

%-----------------------------------------------------------------------------------------------------------------------

\section{Results and Discussion} \label{results}
\begin{figure*}[!t]
\centering 
\subfloat[PSNR: 51.3914]{\includegraphics[width=0.35\columnwidth]{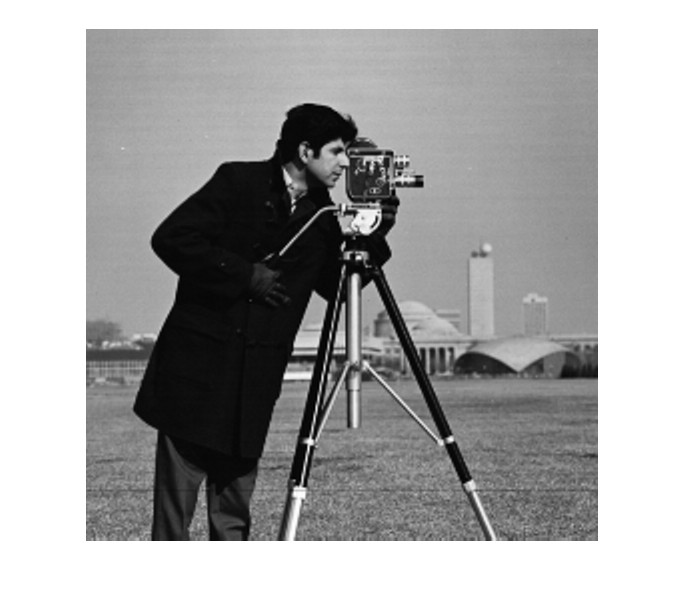}}
\subfloat[PSNR: 50.2963]{\includegraphics[width=0.35\columnwidth]{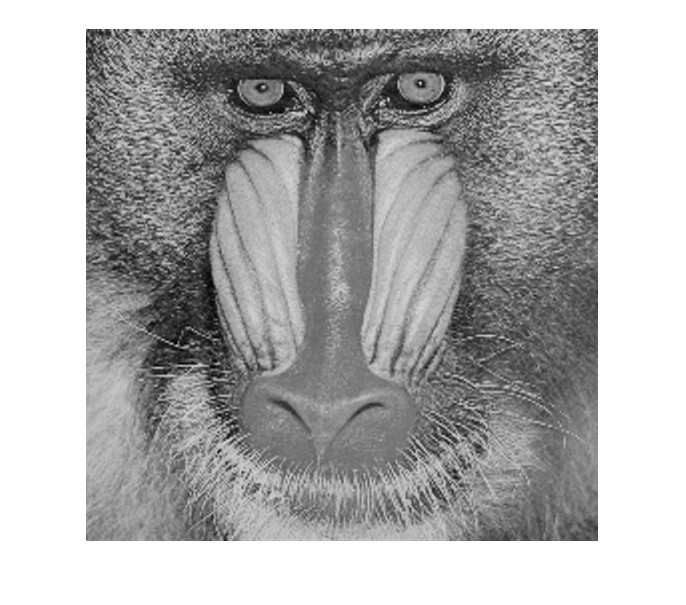}}
\subfloat[PSNR: 51.7377]{\includegraphics[width=0.35\columnwidth]{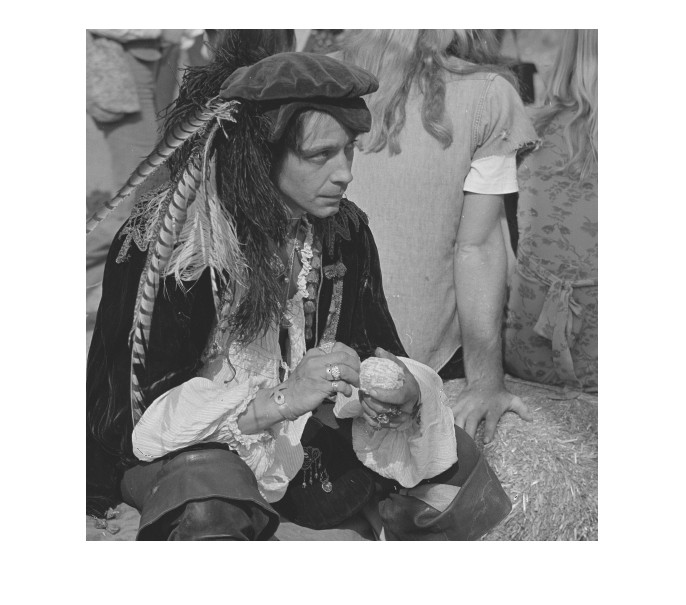}}
\subfloat[PSNR: 51.7131]{\includegraphics[width=0.35\columnwidth]{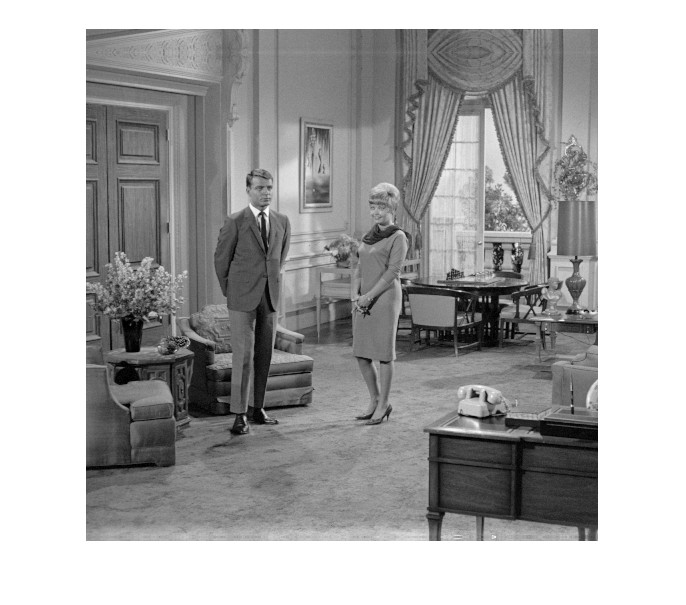}}
\subfloat[PSNR: 51.0706]{\includegraphics[width=0.35\columnwidth]{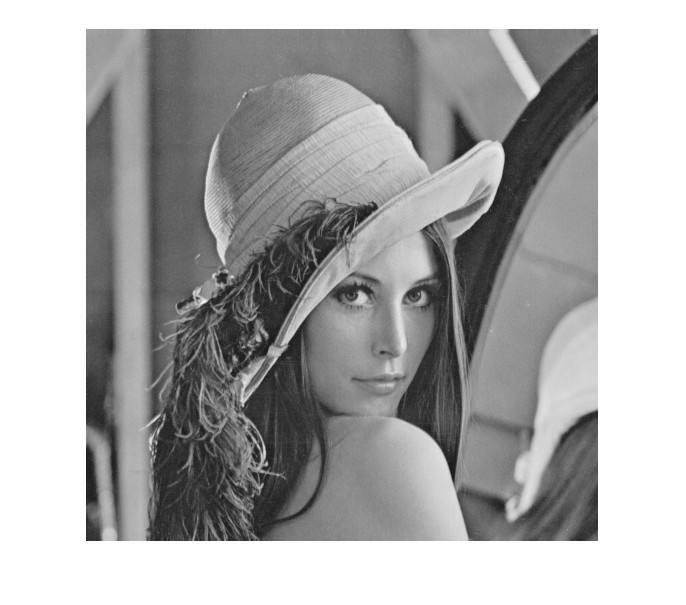}}

\subfloat[PSNR: 38.3342]{\includegraphics[width=0.35\columnwidth]{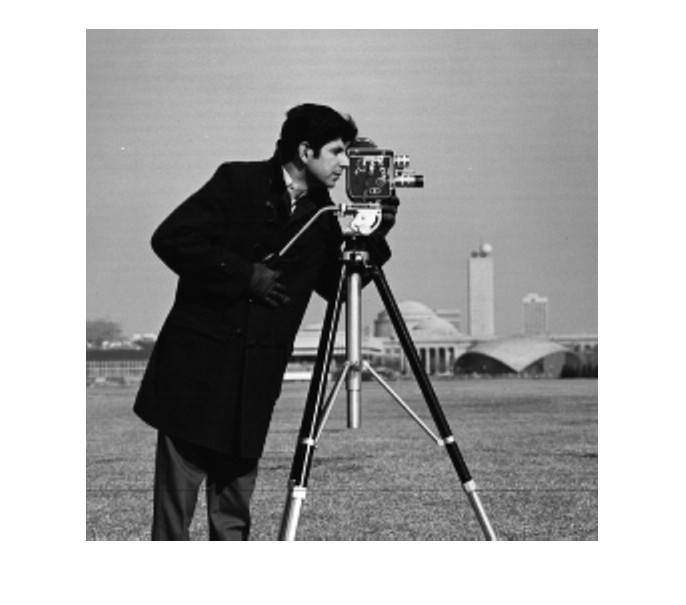}}
\subfloat[PSNR: 36.8322]{\includegraphics[width=0.35\columnwidth]{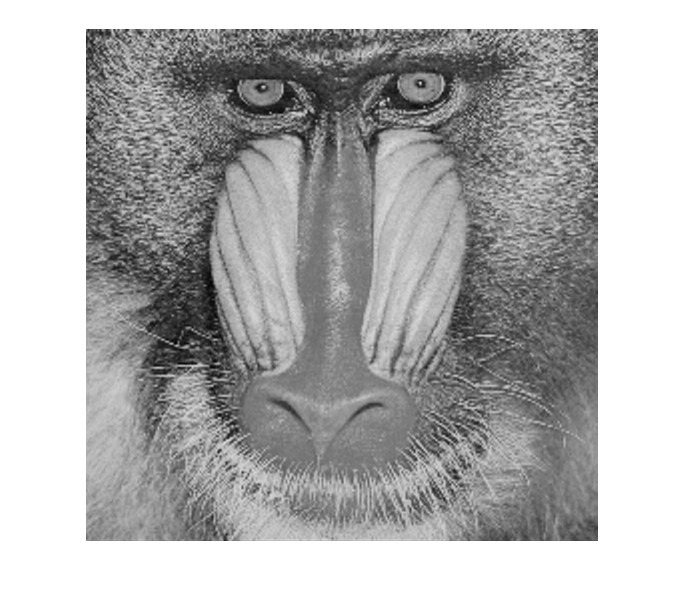}}
\subfloat[PSNR: 38.5621]{\includegraphics[width=0.35\columnwidth]{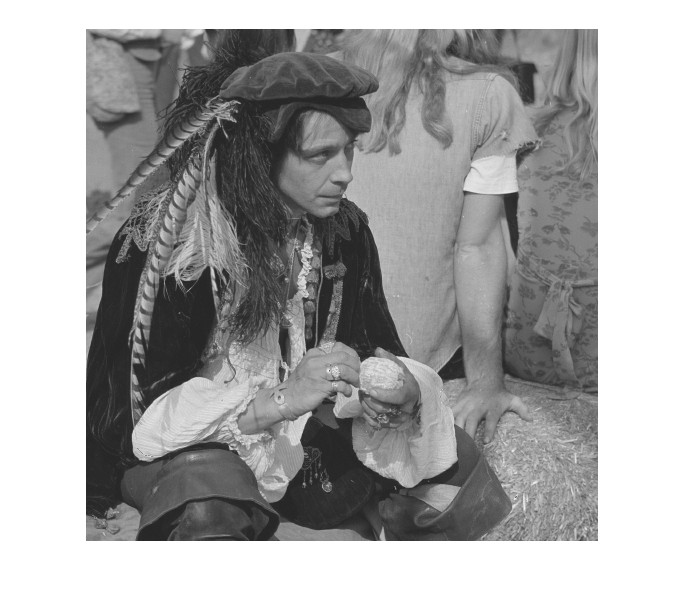}}
\subfloat[PSNR: 38.5924]{\includegraphics[width=0.35\columnwidth]{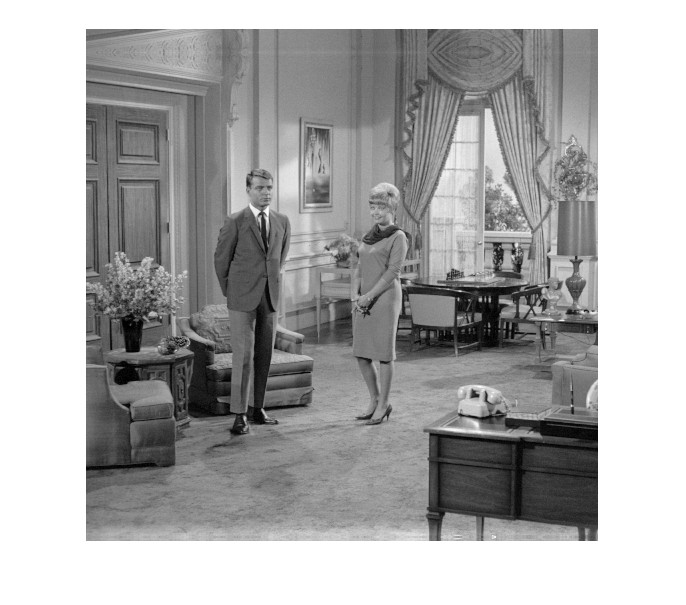}}
\subfloat[PSNR: 38.0782]{\includegraphics[width=0.35\columnwidth]{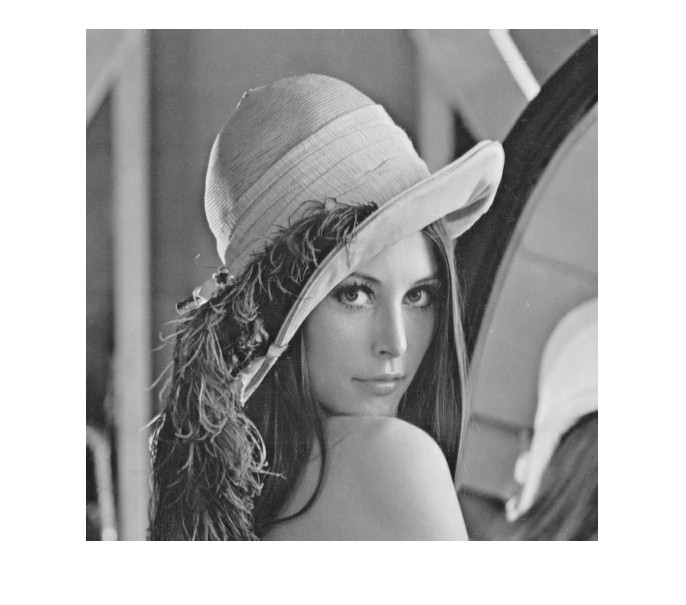}}

\subfloat[PSNR: 53.0773]{\includegraphics[width=0.35\columnwidth]{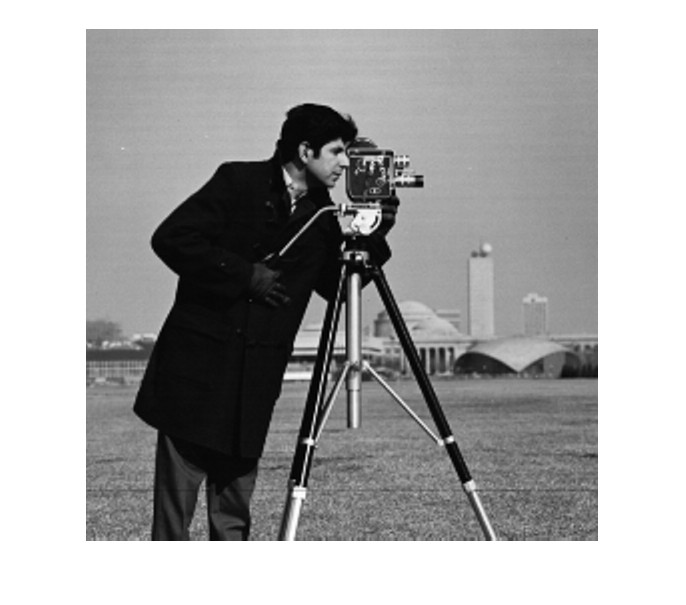}}
\subfloat[PSNR: 50.9789]{\includegraphics[width=0.35\columnwidth]{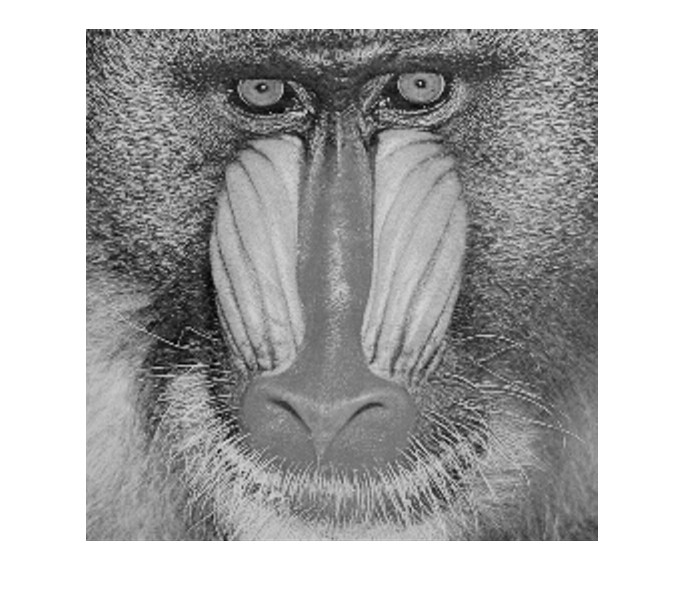}}
\subfloat[PSNR: 53.0270]{\includegraphics[width=0.35\columnwidth]{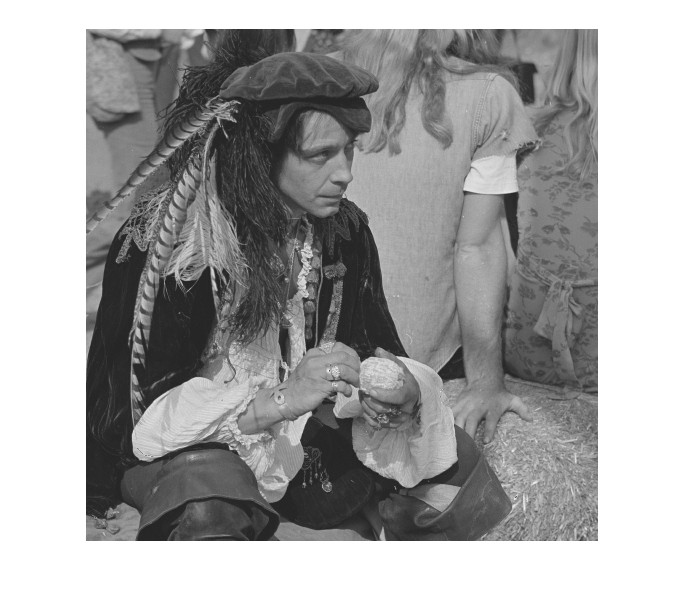}}
\subfloat[PSNR: 53.2469]{\includegraphics[width=0.35\columnwidth]{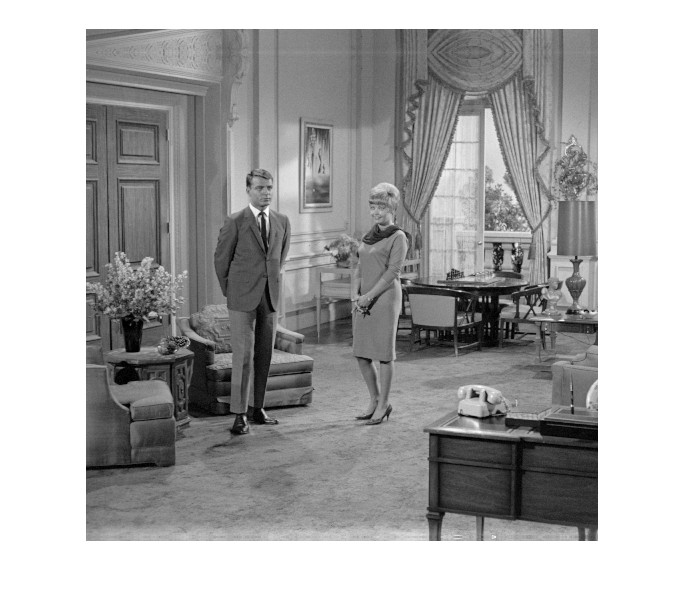}}
\subfloat[PSNR: 53.6129]{\includegraphics[width=0.35\columnwidth]{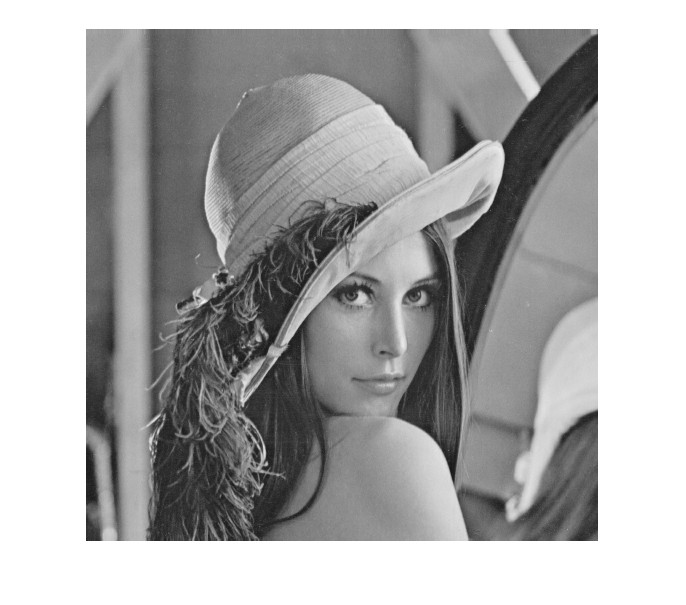}}
\caption{Images obtained after DCT and IDCT using different CORDIC architectures (a)-(e): conventional CORDIC, (f)-(j): lookahead CORDIC and (k)-(o): RICO}
\label{figure:psnrimages}
\end{figure*}

The following sub-sections present the comparison of MATLAB simulation results for the CORDIC architectures discussed in this paper. Somen of the CORDIC architectures have been implemented and their performance comparison has been done. The DCT coefficients are generated using different CORDIC architectures and the errors in those coefficients with respect to the accurate DCT coefficients has been provided in Table \ref{erroranalysis}. The DCT coefficients obtained from conventional, lookahead CORDIC and RICO are used to form individual DCT matrices. These matrices are applied on some test images and the corresponding PSNR values have been obtained. Figure \ref{figure:psnrimages} shows the qualitative analysis for this.

\begin{table}[!h]
\centering
\caption{Error ($\%$) in DCT values}
\begin{tabular}{p{1.5cm}p{2cm}p{2cm}p{1.5cm}}
\hline \noalign {\smallskip}
CORDIC $\longrightarrow$ Coefficient $\downarrow$  & Conventional \cite{volder1959cordic}  &  Lookahead \cite{lee2014reconfigurable} & RICO \cite{Nawandar2016} \\
\noalign {\smallskip} \hline \noalign {\smallskip} 
a & 1.8964  & 3.2422 & 0.387 \\

b & 0.9958 &  0.2164 & 0.2814 \\

c & 5.0276  & 1.1065 & 0.1202 \\

d & 0  & 0.5657 & 0.0282 \\

e & 7.6673 & 2.4118 & 0.503 \\

f & 0.47048 &  26.6596 & 1.5682 \\

g & 6.7692 &  3.1794 & 14.7 \\
\noalign {\smallskip} \hline
\end{tabular}
\label{erroranalysis}
\end{table}

\begin{table}[!h]
\centering
\caption{Quantitative analysis}
\begin{tabular}{p{1.8cm}p{2cm}p{2cm}p{1.5cm}}
\hline \noalign {\smallskip}
CORDIC $\longrightarrow$          & Conventional \cite{volder1959cordic}  & Lookahead \cite{lee2014reconfigurable} & RICO \cite{Nawandar2016} \\
Design metrics $\downarrow$       &     &    &     \\
\noalign {\smallskip} \hline \noalign {\smallskip}
MSE & 0.427  &  9.049 & 0.3296	\\

PSNR & 53.512  & 38.25 & 52.556 \\

SSIM & 1 & 0.9932 & 0.99976 \\

GMSD & 0.000186 & 0.00393 & 0.000724 \\
\noalign {\smallskip} \hline
\end{tabular}
\label{qa}
\end{table}

\section{Conclusion} \label{conclusion}
CORDIC is a class of algorithm which is capable of performing simple to highly 
complex mathematical operations, functions, etc. by the use of adders and shifters only which makes it both hardware and time efficient. Various types of architectures like, the conventional one, a scale-free type of design, lookahead CORDIC i.e. the one employing a concept parallel to that used in a carry lookahead adder (CLA), radix-4 CORDIC, angle recoding, hybrid CORDIC, etc. have been mentioned in the paper. Apart from the survey, these architectures have been used to generate circular trigonometric values for which they are operated in the rotation mode. To solve the purpose MATLAB has been used as it is an easy way to represent the functionality of the architectures and the accuracy of the designs is tested here. Qualitative and quantitative analysis of the mentioned architectures is also done using MATLAB and accordingly the CORDIC architectures have been compared. To verify the efficacy of these architectures, they have been used to obtain the DCT coefficients are found out which is further used to form an 8x8 DCT matrix, to be used to process benchmark images and videos. Finally the architectures are compared on the basis of the design metrics obtained from MATLAB and Design compiler.

%----------------------------------------------------------------------------------------------------------------------

% \section{Future scope} \label{future}
% The RICO architecture can be further modified to a reconfigurable one, where one can adjust the amount of error to be introduced in the design. Also one can get the desired accurate or approximate results by configuring the architecture such that it can be used in different modes (error modes) as per the application requirement.

%----------------------------------------------------------------------------------------------------------------------

\addcontentsline{toc}{chapter}{REFERENCES}
\bibliographystyle{ieeetr}
\bibliography{reference}

% \end{document}

% details about the authors
% \begin{biography}[{\includegraphics[width=1in,height=1.25in,clip,keepaspectratio]{NehaKN.jpeg}}]{Neha K. Nawandar}
% received her B.E. degree in Electronics and Communication from Shri Ramdeobaba Kamla Nehru Engineering College, Nagpur, in the year 2012, and M. Tech. (VLSI Design) from ABV-Indian Institute of Information Technology and Management Gwalior, India, in 2015. She is currently a research scholar at Visvesvaraya National Institute of Technology, Nagpur. Her research interest includes low power and approximate computation architectures for DSP applications, applications for the Internet of Things (IoT).
% \end{biography}
% 
% \begin{biography}
% [{\includegraphics[width=1in,height=1.25in,clip,keepaspectratio]{Sir.png}}]{Vishal R. Satpute}
% working as Assistant Professor in Electronics and Communication Engineering Department at VNIT, Nagpur, received his Bachelor's degree in Engineering from Nagpur University in the year 2001 in the field of Electronics Engineering, and Master's degree M. Tech. (Communication Systems) from Indian Institute of Technology Madras in 2003. He completed his Ph.D. (Electronics and Communication) from Visvesvaraya National Institute of Technology Nagpur in 2015 and has a teaching experience of 13 years.
% \end{biography}

\end{document}